\documentclass[conference]{IEEEtran}
\IEEEoverridecommandlockouts
\usepackage{cite}
\usepackage{amsmath,amssymb,amsfonts}
\usepackage{graphicx}
\usepackage{textcomp}
\usepackage{xcolor}
\usepackage{balance}
\usepackage{cite}
\usepackage{hyperref}
\usepackage{amsmath,amssymb,amsfonts}
\usepackage{amsmath}
\usepackage{amsthm}

\usepackage{graphicx}
\usepackage{textcomp}
\usepackage{xcolor}
\usepackage{graphicx}
\usepackage{float}
\usepackage{subfigure}
\usepackage{amsmath}
\usepackage{amsfonts,amssymb}
\usepackage{mathrsfs}
\usepackage{mathtools}
\usepackage{bm}
\usepackage{multirow}
\usepackage{array}
\usepackage{amssymb}
\usepackage{amsmath}
\usepackage{cite}
\usepackage{url}
\usepackage{xcolor}
\usepackage{cite,graphicx,amsmath,amssymb}
\usepackage{subfigure}
\usepackage{fancyhdr}
\usepackage{mdwmath}
\usepackage{mdwtab}
\usepackage{caption}
\usepackage{amsthm}
\usepackage{setspace}
\usepackage{bm}
\usepackage{mathtools}
\usepackage{dsfont}
\usepackage{bbm}
\usepackage{algorithm}
\usepackage{algorithmic}

\newtheorem{theorem}{Theorem}

\newtheorem{lemma}{Lemma}

\newtheorem{corollary}{Corollary}

\makeatletter
\newcommand{\biggg}{\bBigg@{3}}
\newcommand{\Biggg}{\bBigg@{3.5}}
\makeatother
\def\BibTeX{{\rm B\kern-.05em{\sc i\kern-.025em b}\kern-.08em
    T\kern-.1667em\lower.7ex\hbox{E}\kern-.125emX}}
\begin{document}

\title{Lens Antenna Arrays-Assisted mmWave MU-MIMO Uplink Transmission: Joint Beam Selection and Phase-Only Beamforming Design}

\author{\IEEEauthorblockN{Chongjun Ouyang, Hao Xu, Xujie Zang, and Hongwen Yang}
School of Information and Communication Engineering\\
Beijing University of Posts and Telecommunications, Beijing, 100876, China\\
\{DragonAim, Xu\_Hao, zangxj, yanghong\}@bupt.edu.cn}

\maketitle

\begin{abstract}
This paper considers a lens antenna array-assisted millimeter wave (mmWave) multiuser multiple-input multiple-output (MU-MIMO) system. The base station’s beam selection matrix and user terminals’ phase-only beamformers are jointly designed with the aim of maximizing the uplink sum rate. In order to deal with the formulated mixed-integer optimization problem, a penalty dual decomposition (PDD)-based iterative algorithm is developed via capitalizing on the weighted minimum mean square error (WMMSE), block coordinate descent (BCD), and minorization-maximization (MM) techniques.  Moreover, a low-complexity sequential optimization (SO)-based algorithm is proposed at the cost of a slight sum rate performance loss. Numerical results demonstrate that the proposed methods can achieve higher sum rates than state-of-the–art methods.
\end{abstract}

\begin{IEEEkeywords}
Beam selection, millimeter wave, multiuser multiple-input multiple-output, phase-only beamforming.
\end{IEEEkeywords}

\section{Introduction}
Millimeter wave (mmWave) massive multiple-input multiple-output (MIMO) is envisioned to dominate the future wireless network market by supporting huge data rate growth and wide bandwidth. Yet, to reap the maximal data transmission rate, we should connect each antenna to a dedicated Radio Frequency (RF) chain, which is very cost-prohibitive and power-hungry \cite{Gao2018}. One possible and very promising approach to address this limitation is to exploit lens antenna arrays (LAAs) \cite{Zhang2020}. Owing to their angle-dependent energy-focusing capabilities, LAAs can transform the spatial mmWave channel into a sparser beamspace channel where only a small number of energy-focused beams carry most of the information. After beam selection, the number of required RF chains can be reduced with a negligible data rate loss \cite{Gao2018}.

Within this context, a critical problem in LAA-assisted mmWave MIMO systems is the design of effective beam selection schemes. Recently, attentions have been paid to focusing on the design of beam selection and associated digital/analog beamforming. For example, the authors in \cite{Gao2016} proposed an interference aware (IA)-based beam selection algorithm that can achieve near optimal sum rate performance under the zero-forcing (ZF) precoding. The contribution in \cite{Tataria2018,Guo2018,Feng2019,Liu2021} further extended this IA-based method to other scenarios in order to improve the system spectral efficiency. The joint beam selection and user scheduling problem was studied in \cite{Cheng2020,Sun2022} with the objective of maximizing the downlink sum rate. In addition, LAA-assisted wideband transmissions were investigated in various literatures such as \cite{Feng2019,Cheng2022}. In a nutshell, these works have laid a solid foundation for understanding the design of beam selection schemes.

Yet, all the aforementioned works assumed that the user terminals (UTs) are single-antenna devices \cite{Gao2016,Tataria2018,Guo2018,Feng2019,Liu2021,Cheng2020,Sun2022,Cheng2022}; hence, the joint design of base station (BS)'s beam selection matrix and UTs' beamformers was not feasible therein. In fact, due to the decreased wavelength in mmWave frequency and therefore reduced antenna size, it makes practical sense to equip each UT with an antenna array for beamforming design. On this condition, one can further improve the system throughput via a proper joint BS's beam selection and UTs' beamforming design. Unfortunately, by now, there has been very limited work on this topic of joint design, while only a paper appeared recently \cite{Cheng2022_2}. Particularly, the authors in \cite{Cheng2022_2} jointly optimized the BS's beam selection matrix and UTs' beamformers for an uplink mmWave multiuser MIMO (MU-MIMO) system. However, the proposed scheme is based on the maximum ratio combining (MRC), which might suffer from significant performance degradation in terms of uplink sum rate. Hence in this work, we consider the joint design of UTs' beamformers and BS's beam selection matrix as well as combining matrix for the sake of improving the sum rate of mmWave uplink MU-MIMO channels. Besides, considering the limitation of size as well as energy consumption at the UT side, we assume that each UT is equipped with a phased antenna array connected with a single RF chain. Under this set-up, the UT can exploit the so-called phase-only beamforming \cite{Lau2014}.

The contributions of this paper are summarized as follows. For a mmWave uplink MU-MIMO channel, we propose a joint BS's beam selection and UTs' phase-only beamforming design algorithm for maximizing the sum rate. The formulated mixed-integer non-convex optimization problem is solved by a penalty dual decomposition (PDD)-based algorithm via capitalizing on the weighted minimum mean square error (WMMSE), block coordinate descent (BCD), and minorization-maximization (MM) techniques. To further reduce the computational complexity, we also propose a sequential optimization (SO)-based algorithm that requires no iterations at the expense of a slight sum rate performance loss. Numerical results verify that the proposed methods can achieve higher sum rates than existing methods.

\section{System Model and Problem Formulation}
\subsection{Phase-Only Beamforming-based Beamspace MIMO}
\begin{figure}[!t]
    \centering
    \subfigbottomskip=0pt
	\subfigcapskip=-5pt
    \setlength{\abovecaptionskip}{0pt}
    \includegraphics[width=0.4\textwidth]{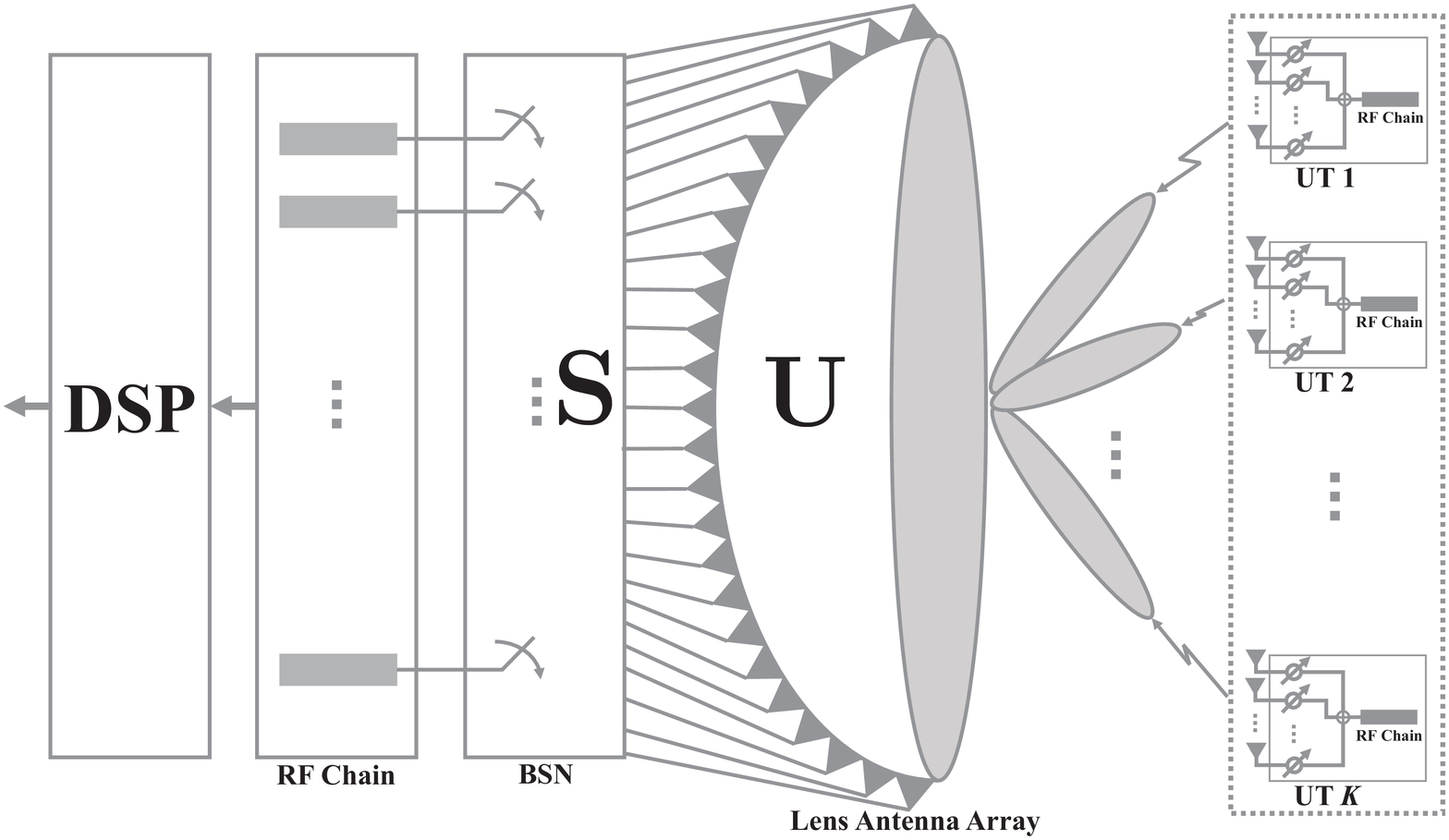}
    \caption{An uplink beamspace MIMO channel with $K$ UTs.}
    \label{system_model}
    \vspace{-20pt}
\end{figure}

The considered mmWave MU-MIMO uplink communication system is depicted in {\figurename} {\ref{system_model}}, consisting of $K$ UTs and one BS. To reduce the hardware implementation costs, we assume that each UT $k\in\mathcal{K}\triangleq\left\{1,2,\ldots,K\right\}$ is equipped with a single RF chain and an $N_k$-element phased array to convey signals while the BS has an $N$-antenna LLA and $L$ ($K\leq L\ll N$) RF chains for receiving. We denote $\mathbf{x}_k\in{\mathbbmss{C}}^{N_{k}\times1}$ as the signal vector sent by UT $k$, which satisfies $\mathbf{x}_k=\mathbf{w}_kx_k$ with $x_k\in{\mathbbmss{C}}$ and $\mathbf{w}_k\in{\mathbbmss{C}}^{N_{k}\times1}$ denoting the data symbol and phase-only beamforming vector, respectively. Particularly, we have $\mathbbmss{E}\left\{x_k\right\}=0$, $\mathbbmss{E}\left\{x_kx_{k'}^{\mathsf{H}}\right\}=0$, $\forall k\neq k'$, and $\mathbf{w}_k=\frac{1}{\sqrt{N_k}}{\bm\phi}_k$, where ${\bm\phi}_k=\left[\phi_{k,1};\ldots;\phi_{k,N_k}\right]$, $\phi_{k,n_k}={\rm{e}}^{{\rm{j}}\theta_{{n_k}}}$, $n_k\in\left\{1,\ldots,N_k\right\}$, with ${\rm{j}}^2=-1$ and $\theta_{n_k}\in\left[0,2\pi\right)$ being the phase shift introduced by the $n$th element of the phased array at UT $k$. Then, the received signal at the BS is
\begin{align}
{\mathbf{y}}=\sum\nolimits_{k=1}^{K}\mathbf{H}_k\mathbf{w}_kx_k+\mathbf{n},
\end{align}
where $\mathbf{H}_k\in{\mathbbmss{C}}^{N\times N_k}$ represents the channel matrix from UT $k$ to the BS, and $\mathbf{n}\sim{\mathcal{CN}}\left(\mathbf{0},\sigma^2\mathbf{I}_N\right)$ is the thermal noise at the BS with $\sigma^2$ being the noise power.

After passing through the LLA and the beam selection network (BSN), the received signal vector becomes \cite{Gao2016,Tataria2018,Guo2018,Feng2019,Liu2021,Cheng2020,Sun2022}
\begin{align}
{\mathbf q}={\mathbf S}{\mathbf U}{\mathbf y}={\mathbf S}{\mathbf U}\sum\nolimits_{k=1}^{K}{\mathbf H}_k{\mathbf w}_kx_k+{\mathbf S}{\mathbf U}{\mathbf n},
\end{align}
where ${\mathbf U}\in{\mathbbmss C}^{N\times N}$ is a discrete Fourier transform (DFT) matrix that reflects the influence of LLA on the incident signal, and ${\mathbf S}=\left[S_{l,n}\right]\in{\mathbbmss C}^{L\times N}$ is the beam selection matrix with binary entries $S_{l,n}\in\left\{0,1\right\}$ to choose beams. It is worth noting that $S_{l,n}=1$ represents the $n$th beam is selected to connect with the $l$th RF chain. In general, each beam is selected for at most one RF chain, which yields ${\mathbf{S}}{\mathbf{S}}^{{\mathsf{H}}}={\mathbf{I}}_L$ \cite{Guo2018}. Afterward, the signal vector $\mathbf{q}$ will be sent to the digital signal processor (DSP) for symbol detection as well as information decoding.

As is widely know, the sum rate capacity of an uplink MU-MIMO channel can be achieved by exploiting a minimum mean square error with successive interference cancellation (MMSE-SIC) decoder. Particularly, the uplink sum rate achieved by the MMSE-SIC decoder is given by
\begin{align}\label{sum_rate_function}
\mathcal{R}=\log_2\!\det\!\left(\!{\mathbf{I}}_L+\sum\nolimits_{k=1}^{K}\frac{p_k}{\sigma^2}{\mathbf S}{\mathbf U}{\mathbf H}_k{\mathbf w}_k{\mathbf w}_k^{\mathsf{H}}
{\mathbf H}_k^{\mathsf{H}}{\mathbf U}^{\mathsf{H}}{\mathbf S}^{\mathsf{H}}\!\right)
\end{align}
with $p_k=\mathbbmss{E}\{|x_k|^2\}$ denoting the transmit power of UT $k$.

\subsection{mmWave Channel Model}
The mmWave spatial channel can be characterized by the famous Saleh-Valenzuela (S-V) model. By assuming that the BS and all the UTs are equipped with uniform linear arrays (ULAs), the spatial channel of UT $k$ can be written as
\begin{align}
{\mathbf{H}}_k=\sqrt{\frac{{\mu}_k}{M_k}}\sum\nolimits_{l=1}^{M_k}\beta_{k,l}{\mathbf{a}}(\varphi_{k,l}){\mathbf{a}}_{N_{k}}^{{\mathsf{H}}}(\theta_{k,l})
\end{align}
where $M_k$ denotes the total number of resolvable paths, $\mu_k$ is the large-scale fading gain between the BS and UT $k$, $\beta_{k,l}\sim{\mathcal{CN}}\left(0,1\right)$ denotes the complex gain of the $l$th path, ${\mathbf{a}}_T(\psi)$ is the array response of a $T$-element ULA with $\psi$ being the angle of arrival (AoA) or angle of departure (AoD), $\varphi_{k,l}$ denotes the AoA at the BS, and $\theta_{k,l}$ denotes the AoD at UT $k$. Particularly, we have ${\mathbf{a}}_T(\psi)=[{\rm e}^{-{\rm j}2\pi\frac{d}{\lambda}x\sin\psi}]_{x\in{\mathcal{I}(T)}}$, where $\mathcal{I}(T)=\left\{-\frac{T-1}{2}, -\frac{T+1}{2},\ldots, \frac{T-1}{2}\right\}$ is a symmetric set of indices centered around zero, $\lambda$ denotes the wavelength, and $d$ denotes the inter-antenna space. Here, we set $d=\frac{\lambda}{2}$.

\subsection{Problem Formulation}
In this work, we investigate the transmission strategy for the LLA-assisted MU-MIMO uplink system, where we strive to jointly optimize the transmit power, $p_k$, $\forall k$, as well as the phase-only beamforming vector $\mathbf{w}_k$, $\forall k$, at the UT sides and the beam selection matrix, $\mathbf{S}$, for improving the system performance. From a systematic perspective, we adopt the sum rate of the entire communication system as our design criterion. Then, the optimization of $p_k$, $\mathbf{w}_k$, $\forall k$, and $\mathbf{S}$ is formulated as the following problem
\begin{subequations}
\label{P_1}
\begin{align}
\mathcal P_1:\max_{{\mathbf S},{ p},{\mathbf{w}}}~&{\mathcal R}\label{p0_OF}\\
{\rm{s.t.}}~&S_{l,n}\in\left\{0,1\right\},l=1,\cdots,L,n=1,\ldots,N,\label{p0_cons1}\\
&{\mathbf{S}}{\mathbf{S}}^{{\mathsf{H}}}={\mathbf{I}}_L,0\leq p_k\leq p_{k,\max},\forall k\in{\mathcal{K}},\label{p0_cons2}\\
&|\phi_{k,n_k}|=1,k=1,\ldots,K,n_k=1,\ldots,N_k,\label{p0_cons3}
\end{align}
\end{subequations}
where ${p}\triangleq\left\{p_k\right\}_{k=1}^{K}$, $\mathbf{w}\triangleq\left\{{\mathbf{w}}_k\right\}_{k=1}^{K}$, and $p_{k,\max}$ is the maximum available
transmit power at UT $k$. It is worth mentioning that the optimization problem $\mathcal P_1$ in \eqref{P_1} presents a different mathematical form from that defined in \cite[Equ. (2)]{Cheng2022_2}; therefore, the proposed methods in \cite{Cheng2022_2} are not applicable to problem $\mathcal P_1$. It is also worth noting that problem \eqref{P_1} is quite challenging due to the following reasons. First, the objective function of $\mathcal P_1$ in \eqref{p0_OF} presents a complicated form. Second, the discrete constraints in \eqref{p0_cons1} make $\mathcal P_1$ essentially an NP-hard problem. Additionally, the presence of the unit-modulus constrained
${\bm\phi}\triangleq\left[{\bm\phi}_1^{\mathsf{H}},\ldots,{\bm\phi}_{K}^{\mathsf{H}}\right]^{\mathsf{H}}\in{\mathbbmss{C}^{\sum_{k=1}^{N_k}\times1}}$ further complicates the optimization procedure. In the sequel, we intend to formulate an efficient approach to address this difficult mixed-integer non-convex problem.

\section{PDD-Based Joint Optimization Algorithm}
It is worth mentioning that $\mathcal{R}$ in \eqref{sum_rate_function} can be written as
\begin{align}\label{System_Model_Simplify_1}
\mathcal{R}=\log_2\det\left(\mathbf{I}_L+{\sigma^{-2}}{\mathbf{S}}{\mathbf{H}}{\bm\Lambda}{\mathbf{H}}^{\mathsf{H}}{\mathbf{S}}^{\mathsf{H}}\right),
\end{align}
where ${\mathbf{H}}=\left[{\mathbf{U}}{\mathbf H}_1{\mathbf w}_1,\ldots,{\mathbf{U}}{\mathbf H}_K{\mathbf w}_K\right]\in{\mathbbmss{C}}^{N\times K}$ and ${\bm\Lambda}=\mathsf{diag}\left\{p_1,\ldots,p_K\right\}\in{\mathbbmss{C}}^{K\times K}$. Particularly, we have
\begin{align}
\frac{\partial{\mathcal{R}}}{\partial p_k}=\frac{1}{\sigma^2}\!
\left[{\mathbf{H}}^{\mathsf{H}}{\mathbf{S}}^{\mathsf{H}}\left(\mathbf{I}_L+\frac{1}{\sigma^2}{\mathbf{S}}{\mathbf{H}}
{\bm\Lambda}{\mathbf{H}}^{\mathsf{H}}{\mathbf{S}}^{\mathsf{H}}\right)^{-1}{\mathbf{S}}{\mathbf{H}}\right]_{k,k}.
\end{align}
It is clear that $\frac{\partial{\mathcal{R}}}{\partial p_k}\geq0$, which suggests that $\mathcal{R}$ monotonically increases with $p_k$. Consequently, it is always desirable to transmit at the maximum power, namely the optimal solution to $p_k$ is $p_k=p_{k,\max}$, $\forall k\in{\mathcal{K}}$. Moreover, by exploiting the structure of the beam selection matrix $\mathbf{S}$, we have
\begin{align}
{\mathbf{S}}^{\mathsf{H}}{\mathbf{S}}={\mathsf{diag}}\left\{s_1,\ldots,s_N\right\},
\end{align}
where $s_n\in\{0,1\}$, $\forall n=1,\ldots,N$, with $s_n$ equalling 1/0 representing the $n$th beam is/is not selected. Define ${\bm\Delta}\triangleq{\mathbf{S}}^{\mathsf{H}}{\mathbf{S}}$. Then, it has ${\bm\Delta}{\bm\Delta}^{\mathsf{H}}={\mathsf{diag}}\left\{s_1,\ldots,s_N\right\}$. It follows that
\begin{align}\label{Sum_Rate_Transform}
\mathcal{R}=\log_2\det\left(\mathbf{I}_N+{\sigma^{-2}}{\bm{\Delta}}\bar{\mathbf{H}}\bar{\mathbf{H}}^{\mathsf{H}}{\bm{\Delta}}^{\mathsf{H}}\right),
\end{align}
where $\bar{\mathbf{H}}=\left[\sqrt{p_{1,\max}}{\mathbf{U}}{\mathbf H}_1{\mathbf w}_1,\ldots,\sqrt{p_{K,\max}}{\mathbf{U}}{\mathbf H}_K{\mathbf w}_K\right]$ and \eqref{Sum_Rate_Transform} is derived by using the Sylvester's determinant identity. Therefore, problem $\mathcal P_1$ can be simplified as
\begin{subequations}
\label{P_2}
\begin{align}
\mathcal P_2:\max_{{\mathbf s},{\bm\phi}}~&{\mathcal R}=\log_2\det\left(\mathbf{I}_N+{\sigma^{-2}}{\bm{\Delta}}\bar{\mathbf{H}}\bar{\mathbf{H}}^{\mathsf{H}}{\bm{\Delta}}^{\mathsf{H}}\right)\label{p2_OF}\\
{\rm{s.t.}}~&s_n\in\{0,1\},n=1,\ldots,N,\sum\nolimits_{n=1}^{N}s_n=L,\label{p2_cons1}\\
&|\phi_{k,n_k}|=1,k=1,\ldots,K,n_k=1,\ldots,N_k,\label{p2_cons2}
\end{align}
\end{subequations}
where ${\mathbf{s}}=[s_1;\ldots;s_N]$. We then simplify problem $\mathcal P_2$ to a more tractable yet equivalent form based on the framework of WMMSE. On this basis, we can handle the simplified problem based on the PDD method.
\subsection{Weighted Mean Square Error (MSE) Minimization}
To proceed, we regard the sum rate in \eqref{Sum_Rate_Transform} as the data rate of a hypothetical communication system described by ${\mathbf y}_{\rm{h}}={\bm{\Delta}}\bar{\mathbf{H}}{\mathbf s}_{\rm h}+{\mathbf n}_{\rm h}$ with ${\bm{\Delta}}\bar{\mathbf{H}}$, ${\mathbf s}_{\rm h}\sim{\mathcal{CN}}\left({\mathbf 0},{\mathbf I}_{K}\right)$, and ${\mathbf n}_{\rm h}\sim{\mathcal{CN}}\left({\mathbf 0},\sigma^2{\mathbf I}_{N}\right)$ being the hypothetical channel matrix, signal symbol vector, and additive noise, respectively. Assume a matrix ${\mathbf U}_{\rm{h}}\in{\mathbbmss C}^{N\times K}$ is adopted for linear receiving, and thus the MSE matrix is given as follows
\begin{align}
{\mathbf E}_{\rm{h}}&={\mathbbmss{E}}_{{\mathbf s}_{\rm h},{\mathbf n}_{\rm h}}
\left\{\left({\mathbf U}_{\rm{h}}^{\mathsf{H}}{\mathbf y}_{\rm{h}}-{\mathbf s}_{\rm h}\right)\left({\mathbf U}_{\rm{h}}^{\mathsf{H}}{\mathbf y}_{\rm{h}}-{\mathbf s}_{\rm h}\right)^{\mathsf{H}}\right\}\\
&=\left({\mathbf U}_{\rm{h}}^{\mathsf{H}}{\bm{\Delta}}\bar{\mathbf{H}}-{\mathbf I}_{K}\right)\left({\mathbf U}_{\rm{h}}^{\mathsf{H}}{\bm{\Delta}}\bar{\mathbf{H}}-{\mathbf I}_{K}\right)^{\mathsf{H}}+\sigma^2{\mathbf U}_{\rm{h}}^{\mathsf{H}}{\mathbf U}_{\rm{h}}.\label{MSEMatrix}
\end{align}
Then, by the theory of the WMMSE method \cite[Theorem 1]{Shi2011}, we introduce an auxiliary optimization matrix variable ${\mathbf{W}}_{\rm{h}}\in{\mathbbmss{C}}^{K\times K}$ and apply it to establish a matrix-weighted MSE minimization problem as
\begin{subequations}
\label{P_3}
\begin{align}
\mathcal P_3:&\min_{{\mathbf W}_{\rm{h}},{\mathbf U}_{\rm{h}},{\mathbf s},{\bm\phi}}~{\mathsf{tr}}\left\{{\mathbf W}_{\rm{h}}{\mathbf E}_{\rm{h}}\right\}-\log\det\left({\mathbf W}_{\rm{h}}\right)\label{p3_OF}\\
&~{\rm{s.t.}}~s_n\in\{0,1\},n=1,\ldots,N,\sum\nolimits_{n=1}^{N}s_n=L,\label{p3_cons1}\\
&\qquad|\phi_{k,n_k}|=1,k=1,\ldots,K,n_k=1,\ldots,N_k.\label{p3_cons2}
\end{align}
\end{subequations}
Problem $\mathcal P_3$ is equivalent to problem $\mathcal P_2$ in the sense that the solutions of $\left\{{\mathbf s},{\bm\phi}\right\}$ for the two problems are identical \cite[Theorem 1]{Shi2011}. It is worth mentioning that problem $\mathcal P_3$ is more tractable than problem $\mathcal P_2$ as well as problem $\mathcal P_1$ since the objective function is convex over each variable $({\mathbf W}_{\rm{h}},{\mathbf U}_{\rm{h}},{\mathbf s},{\bm\phi})$ while holding others fixed. Furthermore, to deal with the discrete constrains \eqref{p3_cons1}, we introduce the auxiliary variables $\bar{\mathbf{s}}=[\bar{s}_1;\ldots;\bar{s}_N]$ subject to the constraints of ${\bar s}_n=s_n$, $s_n\left(1-{\bar s}_n\right)=0$, and $0\leq {\bar s}_n\leq 1$. Then, we can equivalently convert problem $\mathcal P_3$ as follows:
\begin{subequations}
\label{P_4}
\begin{align}
\mathcal P_4:&\min_{{\mathbf W}_{\rm{h}},{\mathbf U}_{\rm{h}},\bar{\mathbf s},{\mathbf s},{\bm\phi}}~{\mathsf{tr}}\left\{{\mathbf W}_{\rm{h}}{\mathbf E}_{\rm{h}}\right\}-\log\det\left({\mathbf W}_{\rm{h}}\right)\label{p4_OF}\\
&~{\rm{s.t.}}~\bar{s}_n\in\{0,1\},n=1,\ldots,N,\sum\nolimits_{n=1}^{N}s_n=L,\label{p4_cons1}\\
&\qquad|\phi_{k,n_k}|=1,k=1,\ldots,K,n_k=1,\ldots,N_k,\label{p4_cons2}\\
&\qquad{\bar s}_n=s_n,s_n\left(1-{\bar s}_n\right)=0,n=1,\ldots,N.\label{p4_cons3}
\end{align}
\end{subequations}
Particularly, we find that optimizing each variable of problem $\mathcal P_4$ separately while treating the others as constants leads to a series of subproblem that can be easily solved. By exploiting this property, we resort to the PDD technique to formulate a computationally efficient algorithm to tackle problem $\mathcal P_4$.

\subsection{The Proposed PDD-Based Algorithm}
In general, the PDD-based algorithm is characterized by an embedded double loop structure \cite{Shi2020}. Specifically, the inner loop solves the augmented Lagrangian (AL) subproblem while the outer loop aims to update the dual variables or the penalty parameter based on the constraint violation. Then, following the principle of PDD, we convert problem $\mathcal P_4$ into its AL form which is expressed as
\begin{subequations}
\label{P_5}
\begin{align}
\mathcal P_5:&\min_{{\mathbf W}_{\rm{h}},{\mathbf U}_{\rm{h}},\bar{\mathbf s},{\mathbf s},{\bm\phi}}~{\mathsf{tr}}\left\{{\mathbf W}_{\rm{h}}{\mathbf E}_{\rm{h}}\right\}-\log\det\left({\mathbf W}_{\rm{h}}\right)+\frac{f_2({\mathbf{s}},\bar{\mathbf{s}})}{2\rho}\label{p5_OF}\\
&~{\rm{s.t.}}~\bar{s}_n\in\{0,1\},n=1,\ldots,N,\label{p5_cons1}\\
&\qquad|\phi_{k,n_k}|=1,k=1,\ldots,K,n_k=1,\ldots,N_k,\label{p5_cons2}
\end{align}
\end{subequations}
where $\rho>0$ is the penalty parameter,
\begin{align}
f_2({\mathbf{s}},\bar{\mathbf{s}})&=\left(\sum_{n=1}^{N}s_n-L+\rho\xi\right)^2
+\sum_{n=1}^{N}\left[\left(s_n-{\bar s}_n+\rho\mu_n\right)^2\right.\nonumber\\
&\left.+\left(s_n\left(1-{\bar s}_n\right)+\rho\lambda_n\right)^2\right],
\end{align}
and where $\xi$, $\{\mu_n\}_{n=1}^{N}$, and $\{\lambda_n\}_{n=1}^{N}$ denote the dual variables associated with the equality constraints in \eqref{p4_cons1} and \eqref{p4_cons3}. To tackle ${\mathcal P}_{5}$ more conveniently, we resort to the block coordinate descent method to iteratively optimize one block of variables while regarding the others as constants. To be more specific, we minimize the objective function in \eqref{p5_OF} by sequentially updating ${\mathbf W}_{\rm{h}}$, ${\mathbf U}_{\rm{h}}$, $\bar{\mathbf s}$, ${\mathbf s}$, and ${\bm\phi}$ in the inner loop as follows.

In \textbf{Step 1}, we optimize ${\mathbf W}_{\rm{h}}$ by fixing the remaining variables. On this condition, ${\mathcal P}_5$ deduces to the unconstrained
subproblem: ${\mathbf W}_{\rm{h}}^{\star}=\arg\min_{{\mathbf W}_{\rm{h}}}{\mathsf{tr}}\left\{{\mathbf W}_{\rm{h}}{\mathbf E}_{\rm{h}}\right\}-\log\det\left({\mathbf W}_{\rm{h}}\right)$. Using its first-order optimality condition, we get ${\mathbf W}_{\rm{h}}^{\star}={\mathbf E}_{\rm{h}}^{-1}$.

In \textbf{Step 2}, we optimize ${\mathbf U}_{\rm{h}}$ by fixing the remaining variables. Similar to the derivation of ${\mathbf W}_{\rm{h}}^{\star}$, the optimal ${\mathbf U}_{\rm{h}}$ is derived as ${\mathbf U}_{\rm{h}}^{\star}=\left(\sigma^2{\mathbf{I}}_N+{\bm{\Delta}}\bar{\mathbf{H}}{\bar{\mathbf{H}}}^{\mathsf{H}}{\bm\Delta}^{\mathsf{H}}\right)^{-1}{\bm{\Delta}}\bar{\mathbf{H}}$.

In \textbf{Step 3}, we optimize $\bar{\mathbf{s}}=[\bar{s}_1;\ldots;\bar{s}_N]$ by fixing the remaining variables. The subproblem of optimizing $\{\bar{s}_n\}_{n=1}^{N}$ in parallel can be expressed as
\begin{equation}\label{Problem_Auxiliary}
\bar{s}_n^{\star}=\arg\min\nolimits_{\bar{s}_n\in\{0,1\}}\left(a_n\bar{s}_n^2-2b_n\bar{s}_n\right),
\end{equation}
where $a_n=1+s_n^2$ and $b_n=s_n+\rho\lambda_n+s_n^2+s_n\rho\mu_n$. Equation \eqref{Problem_Auxiliary} features a scalar quadratic objective function and its optimal solution is given as follows:
\begin{align}\label{Auxiliary_Optimal}
{\bar{s}}_n^{\star}=
\begin{cases}
0,& {{b_n}/{a_n}<0}\\
{b_n}/{a_n},& {0\leq {b_n}/{a_n}\leq1}\\
1,& {{b_n}/{a_n}>1}
\end{cases}.
\end{align}

In \textbf{Step 4}, we optimize ${\mathbf{s}}=[s_1;\ldots;s_N]$ by fixing the remaining variables. This subproblem can be formulated as
\begin{equation}\label{Beam_Selection_Matrix_Opt}
{\mathbf s}^{\star}=\arg\min\nolimits_{\mathbf s}\left({\mathbf s}^{\mathsf{T}}{\mathbf G}{\mathbf s}-{\mathbf s}^{\mathsf{T}}{\mathbf g}\right),
\end{equation}
where
\begin{subequations}
\begin{align}
{\mathbf G}&\triangleq
\Re\left\{\left({\mathbf U}_{\rm{h}}{\mathbf W}_{\rm{h}}{\mathbf U}_{\rm{h}}^{\mathsf{H}}\right)\odot\left({\mathbf{H}}{\mathbf{H}}^{\mathsf{H}}\right)\right\}+\frac{1}{2\rho}\left({\mathbf 1}{\mathbf 1}^{\mathsf{T}}+{\mathbf I}_N\right)\nonumber\\
&+\frac{1}{2\rho}{{\mathsf{diag}}\left\{\left(1-{\bar s}_1\right)^2,\cdots,\left(1-{\bar s}_N\right)^2\right\}},\nonumber\\
{\mathbf g}&\triangleq2\Re\left\{{\mathbf q}\right\}-\frac{1}{\rho}\left(\left(\rho\beta-L\right){\mathbf 1}+\left(\rho{\bm\lambda}-\bar{\mathbf s}\right)+\rho\left({\mathbf{1}}-\bar{\mathbf s}\right)\odot{\bm\mu}\right),\nonumber\\
{\mathbf q}&\triangleq\left[\left[{\mathbf U}_{\rm{h}}{\mathbf W}_{\rm{h}}{\mathbf{H}}^{\mathsf{H}}\right]_{1,1},\cdots,\left[{\mathbf U}_{\rm{h}}{\mathbf W}_{\rm{h}}{\mathbf{H}}^{\mathsf{H}}\right]_{N,N}\right]^{\mathsf{T}},\nonumber
\end{align}
\end{subequations}
with $[\mathbf{A}]_{i,j}$ denotes the $(i,j)$th element of $\mathbf{A}$, ${\bm\lambda}=[\lambda_1;\ldots;\lambda_N]$, and ${\bm\mu}=[\mu_1;\ldots;\mu_N]$. According to the Schur product theorem, we have $\left({\mathbf U}_{\rm{h}}{\mathbf W}_{\rm{h}}{\mathbf U}_{\rm{h}}^{\mathsf{H}}\right)\odot\left({\mathbf{H}}{\mathbf{H}}^{\mathsf{H}}\right)\succeq{\mathbf 0}$ and thus
${\mathbf G}\succ{\mathbf 0}$, which indicates that \eqref{Beam_Selection_Matrix_Opt} is an unconstrained convex problem. By exploiting its first-order optimality condition, we can get ${\mathbf s}^{\star}=\left({\mathbf G}+{\mathbf G}^{\mathsf{T}}\right)^{-1}{\mathbf g}$.

In \textbf{Step 5}, we optimize ${\bm\phi}$ by fixing the remaining variables. The resultant subproblem of optimizing ${\bm\phi}$ is given by
\begin{subequations}
\label{P_6}
\begin{align}
\mathcal P_6:&\min_{{\bm\phi}}~{\bm\phi}^{\mathsf{H}}\mathbf{B}_{\bm\phi}{\bm\phi}-2\Re\{{\bm\phi}^{\mathsf{H}}{\mathbf{b}}_{\bm\phi}\}\label{p6_OF}\\
&~{\rm{s.t.}}~|\phi_{k,n_k}|=1,k=1,\ldots,K,n_k=1,\ldots,N_k,\label{p6_cons1}
\end{align}
\end{subequations}
where $\mathbf{b}_{\bm\phi}\!=\!\left[{\mathbf{b}}_1^{\mathsf{H}},\ldots,{\mathbf{b}}_K^{\mathsf{H}}\right]^{\mathsf{H}}$, $\mathbf{B}_{\bm\phi}\!=\!{\mathsf{BlkDiag}}\left\{{\mathbf{A}}_1,\ldots,{\mathbf{A}}_K\right\}$, ${\mathbf{A}}_k=\frac{p_{k,\max}}{N_k}{\mathbf H}_k^{\mathsf{H}}{\mathbf{U}}^{\mathsf{H}}{\bm{\Delta}}^{\mathsf{H}}{\mathbf{U}}_{\rm{h}}{\mathbf{W}}_{\rm{h}}{\mathbf{U}}_{\rm{h}}^{\mathsf{H}}{\bm{\Delta}}
{\mathbf{U}}{\mathbf H}_k$, and $\mathbf{b}_k=\sqrt{\frac{p_{k,\max}}{N_k}}{\mathbf H}_k^{\mathsf{H}}{\mathbf{U}}^{\mathsf{H}}{\bm{\Delta}}^{\mathsf{H}}{\mathbf{U}}_{\rm{h}}{\mathbf{W}}_{\rm{h}}\left(:,k\right)$. Here, ${\mathbf{W}}_{\rm{h}}\left(:,k\right)$ denotes the $k$th column of matrix ${\mathbf{W}}_{\rm{h}}$. The optimization of $\bm\phi$ in problem $\mathcal{P}_6$ is challenging since the unit-modulus constraints exhibit non-convexity. In the sequel, we resort to the MM technique \cite{Song2016} for the sake of obtaining a suboptimal solution. To proceed, it is necessary to establish an upper bound for the objective function in \eqref{p6_OF}. Based on \cite{Song2016}, an alternative upper bound for ${\bm\phi}^{\mathsf{H}}\mathbf{B}_{\bm\phi}{\bm\phi}-2\Re\{{\bm\phi}^{\mathsf{H}}{\mathbf{b}}_{\bm\phi}\}$ can be established as
\begin{align}
g\left({\bm\phi}|{\bm\phi}^{(i)}\right)&={\bm\phi}^{\mathsf{H}}{\mathbf{L}}{\bm\phi}-
2\Re\left\{{\bm\phi}^{\mathsf{H}}\left({\mathbf{L}}-{\mathbf{G}}\right){\bm\phi}^{(i)}\right\}\nonumber\\
&+\left({\bm\phi}^{(i)}\right)^{\mathsf{H}}\left({\mathbf{L}}-{\mathbf{G}}\right){\bm\phi}^{(i)}-2\Re\{{\bm\phi}^{\mathsf{H}}{\mathbf{b}}_{\bm\phi}\},
\end{align}
where ${\mathbf{L}}=\lambda_{\max}{\mathbf{I}}_{\sum_{k=1}^{K}N_k}$ with $\lambda_{\max}$ representing the maximum eigenvalue of matrix $\mathbf{G}$, $i$ is the iterative index, and ${\bm\phi}^{(i)}$ denotes the minimizer at the $(i-1)$th iteration of the MM procedure. Then, we arrive at the following surrogate subproblems as
\begin{subequations}
\begin{align}
{\mathcal{P}}_{\rm{mm}}^{(i)}~&\min_{{\bm\phi}}~g\left({\bm\phi}|{\bm\phi}^{(i)}\right)\label{p7_OF}\\
{\rm{s.t.}}~&|\phi_{k,n_k}|=1,k=1,\ldots,K,n_k=1,\ldots,N_k.\label{p7_cons1}
\end{align}
\end{subequations}
It is worth noting that ${\bm\phi}^{\mathsf{H}}{\mathbf{L}}{\bm\phi}=\sum_{k=1}^{K}N_k$. By ignoring the terms independent with $\bm\phi$ in \eqref{p7_OF}, we can write the optimal solution to problem ${\mathcal{P}}_{\rm{mm}}^{(i)}$ as
\begin{align}
{\bm\phi}^{(i+1)}={\rm{e}}^{{\rm{j}}\mathsf{arg}\left(\left({\mathbf{L}}-{\mathbf{G}}\right){\bm\phi}^{(i)}+{\mathbf{b}}_{\bm\phi}\right)},
\end{align}
where ${\mathsf{arg}}\left(\cdot\right)$ means the extraction of phase information. The MM-based approach proposed to handle problem \eqref{P_6} is summarized in Algorithm \ref{Algorithm1}. Based on \cite{Song2016}, Algorithm \ref{Algorithm1} is guaranteed to converge to a stationary point of problem \eqref{P_6}.

\begin{algorithm}[htbp]
  \caption{MM-based algorithm for solving problem \eqref{P_6}}
  \label{Algorithm1}
  \begin{algorithmic}[1]
    \STATE Initialize ${\bm\phi}^{(0)}$ and set iteration index $i=0$
    \REPEAT
      \STATE Calculate ${\bm\phi}^{(i+1)}={\rm{e}}^{{\rm{j}}\mathsf{arg}\left(\left({\mathbf{L}}-{\mathbf{G}}\right){\bm\phi}^{(i)}+{\mathbf{b}}_{\bm\phi}\right)}$
      \STATE Update $i\leftarrow i+1$
    \UNTIL{convergence}
  \end{algorithmic}
\end{algorithm}

After investigating the inner loop of the PDD-based method, we turn our attention to discussing its outer loop. In the outer iteration, the constraint violation is calculated as follows:
\begin{equation}\label{Constant_Vio}
\begin{split}
h\triangleq\max_{\forall n}\left\{\left|\sum\nolimits_{n=1}^{N}\!s_n\!-\!L\right|,\left|{\bar s}_n-s_n\right|,\left|s_n\left(1-{\bar s}_n\right)\right|\right\},
\end{split}
\end{equation}
and update the dual variables by the following expressions: \cite[Table \uppercase\expandafter{\romannumeral1}, Line 4]{Shi2020}
\begin{subequations}\label{Dual_Variable_Update}
\begin{align}
\xi^{(t+1)}&=\xi^{(t)}+{\rho}^{-1}\left(\sum\nolimits_{n=1}^{N}s_n-L\right),\\
\mu_n^{(t+1)}&=\mu_n^{(t)}+{\rho}^{-1}\left({\bar s}_n-s_n\right),\\
\lambda_n^{(t+1)}&=\lambda_n^{(t)}+{\rho}^{-1}s_n\left(1-{\bar s}_n\right),
\end{align}
\end{subequations}
where $t$ denotes the outer iteration index.

\subsection{Convergence and Complexity Analyses}
The overall PDD-based algorithm for solving problem \eqref{P_1} is summarized in Algorithm \ref{Algorithm2}. This algorithm is guaranteed to converge to the set of stationary solutions of problem \eqref{P_1} \cite{Shi2020}. After the convergence analysis of the proposed algorithm, we turn our attention to discussing its computational complexity. For clarity, let $I_{\rm{out}}$ and $I_{\rm{in}}$ denote the numbers of iterations in the outer loop and the inner loop, respectively. As previously stated, the per-iteration complexity in the outer loop is mainly originated from the BCD-based updating in the inner loop, whereas the per-iteration complexity in the inner loop is mainly composed of the complexity of updating variables $\left\{{\bm\phi},{\mathbf{s}},\bar{\mathbf{s}},{\mathbf W}_{\rm{h}},{\mathbf U}_{\rm{h}}\right\}$. It is clear to show that the complexity of computing the optimal results of ${\mathbf W}_{\rm{h}}$, ${\mathbf U}_{\rm{h}}$, ${\mathbf{s}}$, and $\bar{\mathbf{s}}$ is evaluated as ${\mathcal O}\left(K^3\right)$, ${\mathcal O}\left(N^3\right)$, ${\mathcal O}\left(N\right)$, and ${\mathcal O}\left(N^3\right)$, respectively. Moreover, the complexity of the MM-based optimization of $\bm\phi$ in Algorithm \ref{Algorithm1} is given by $\mathcal{O}\left(N_{\rm{U}}^3+I_{\rm{mm}}N_{\rm{U}}^2\right)$ with $N_{\rm{U}}=\sum_{k=1}^{K}N_k$ and $I_{\rm{mm}}$ denoting the number of iterations of Algorithm \ref{Algorithm1}. Generally, we have $N>K$ and $N>N_{\rm{U}}$. Hence, the overall complexity of Algorithm \ref{Algorithm2} is estimated as ${\mathcal O}\left(I_{\text{out}}I_{\text{in}}\left(2N^3+I_{\rm{mm}}N_{\rm{U}}^2\right)\right)$, which is in polynomial time.

\begin{algorithm}[htbp]
  \caption{PDD-based algorithm for solving problem \eqref{P_1}}
  \label{Algorithm2}
  \begin{algorithmic}[1]
    \STATE Initialize primary variables $\left\{{\bm\phi},{\mathbf{s}},\bar{\mathbf{s}},{\mathbf W}_{\rm{h}},{\mathbf U}_{\rm{h}}\right\}$, dual variables $\left\{\beta,{\bm\lambda},{\bm\mu}\right\}$, iteration index $t=0$, threshold $\mu$, penalty factor $\rho>0$, and scaling factor $\chi\in(0,1)$
    \REPEAT
    \REPEAT
      \STATE Update $\left\{{\mathbf W}_{\rm{h}},{\mathbf U}_{\rm{h}},\bar{\mathbf{s}},{\mathbf{s}},{\bm\phi}\right\}$ based on \textbf{Steps 1--5}
    \UNTIL{convergence}
    \STATE Calculate the constraint violation $h$ by \eqref{Constant_Vio}
    \IF{$h < \mu$}
	\STATE Update the dual variable by \eqref{Dual_Variable_Update}
	\ELSE
	\STATE Set $\rho = \chi\rho$
	\ENDIF
    \STATE Set $\mu=\chi h$ and $t=t+1$
    \UNTIL{convergence}
  \end{algorithmic}
\end{algorithm}

\section{A Low-Complexity Sequential Optimization}
To reduce the computational complexity of solving problem $\mathcal{P}_1$, we develop a sequential optimization approach in this section. More specifically, we first optimize the phase-only beamforming vector $\bm\phi$ and then optimize the beam selection matrix $\mathbf{S}$ without iteration.

Looking at the sum expression presented in \eqref{System_Model_Simplify_1}, we can get the effective channel gain of UT $k$ as follows:
\begin{align}
\left\|{\mathbf{U}}{\mathbf H}_k{\mathbf w}_k\right\|^2={N_k^{-1}}{\bm{\phi}}_k^{\mathsf{H}}{\mathbf H}_k^{\mathsf{H}}{\mathbf H}_k{\bm{\phi}}_k.
\end{align}
Motivated by this, we propose a heuristic phase-only beamforming design by optimizing the channel gain of each UT in parallel. This problem can be mathematically expressed as
\begin{subequations}
\begin{align}
{\mathcal P}_{\rm{h}}^{k}:&\max\nolimits_{{\bm\phi}_k}~{\bm\phi}_k^{\mathsf{H}}{\mathbf H}_k^{\mathsf{H}}{\mathbf H}_k{\bm\phi}_k\\
&~{\rm{s.t.}}~|\phi_{k,n_k}|=1,n_k=1,\ldots,N_k.
\end{align}
\end{subequations}
Problem ${\mathcal P}_{\rm{h}}^{k}$ can be solved by the MM-based method. To further release the computational burden, we propose an heuristic method by setting ${\bm\phi}_k={\rm{e}}^{\rm{j}\rm{arg}\left(\mathbf{v}_k\right)}$ with $\mathbf{v}_k$ being the principle eigenvector of ${\mathbf H}_k^{\mathsf{H}}{\mathbf H}_k$. Then, the computational complexity involved in the phase shifts design is estimated as $\mathcal{O}\left(N_{\rm{U}}N^2\right)$. We then optimize the beam selection matrix $\mathbf{S}$ after all the phase shifts have been configured. The resultant problem of beam selection is given by
\begin{align}\label{Beam_Selection_Matrix_Opt_Sequential}
\arg\max\nolimits_{\mathbf{S}\in{\mathcal{S}}}\log_2\det\left(\mathbf{I}_L+{\sigma^{-2}}{\mathbf{S}}{\mathbf{H}}{\bm\Lambda}{\mathbf{H}}^{\mathsf{H}}{\mathbf{S}}^{\mathsf{H}}\right),
\end{align}
where $\mathcal{S}$ denotes the candidate set of the feasible beam selection matrices. Problem \eqref{Beam_Selection_Matrix_Opt_Sequential} can be optimally solved via an exhaustive search, which yields exponential search complexity. To reduce the search complexity, we leverage a greedy search-based method to find a high-quality solution to problem \eqref{Beam_Selection_Matrix_Opt_Sequential}. More details about this method can be found in \cite{Gershman2004} and are omitted here due to page limitations. With the aid of \cite{Gershman2004}, we find the computational complexity of the sequential optimization can be estimated as $\mathcal{O}\left(N_{\rm{U}}N^2+LKN+LN\right)$. In the considered system, we have $N\geq L\geq K$ and $N\geq N_{\rm{U}}$, and thus the sequential optimization-based method involves lower complexity than the PDD-based method.

\begin{figure}[!t]
    \centering
    \subfigbottomskip=0pt
	\subfigcapskip=-5pt
\setlength{\abovecaptionskip}{0pt}
    \subfigure[Convergence of sum rate.]
    {
        \includegraphics[height=0.185\textwidth]{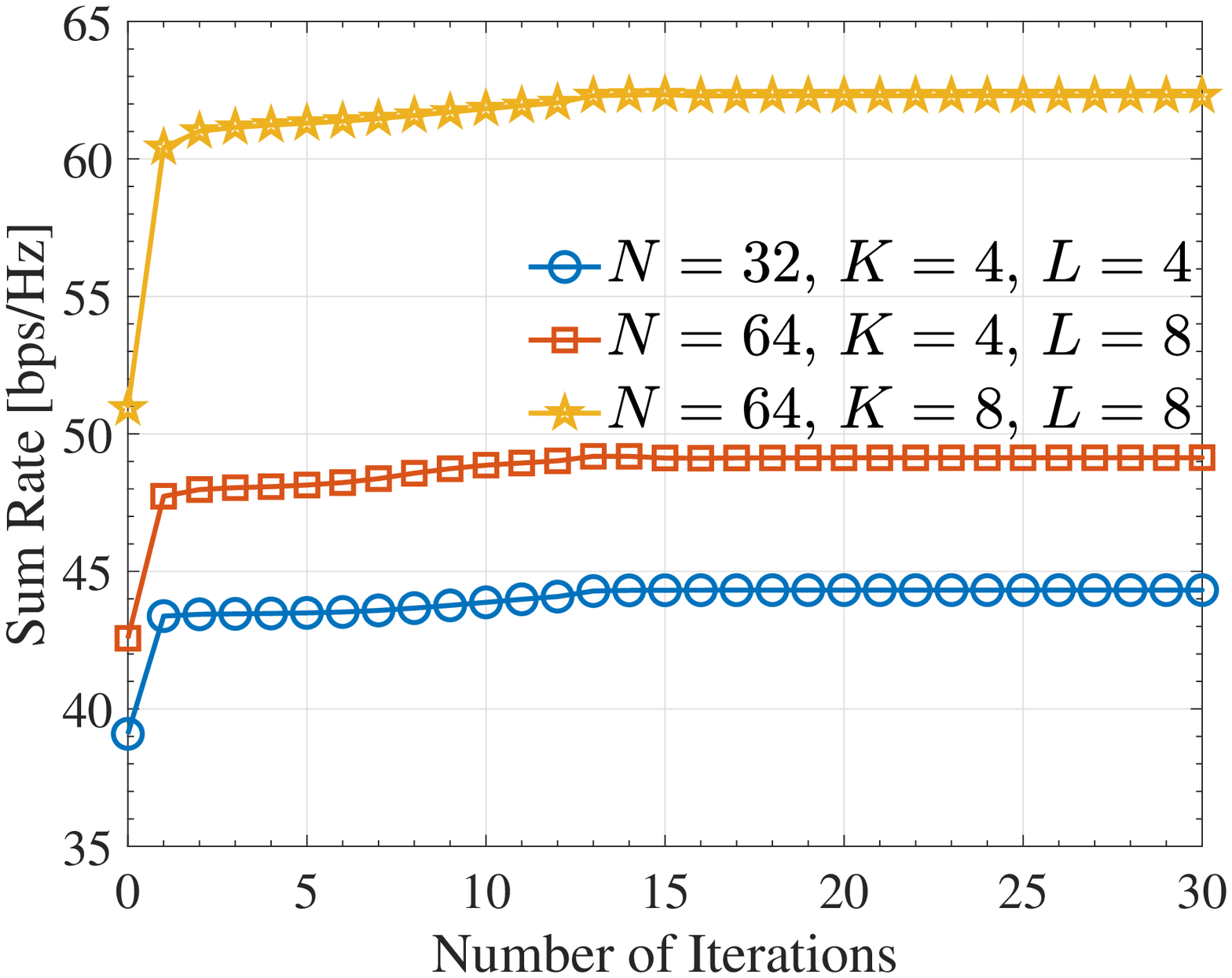}
	   \label{fig1a}	
    }\hspace{-10pt}
   \subfigure[Constraint violation.]
    {
        \includegraphics[height=0.185\textwidth]{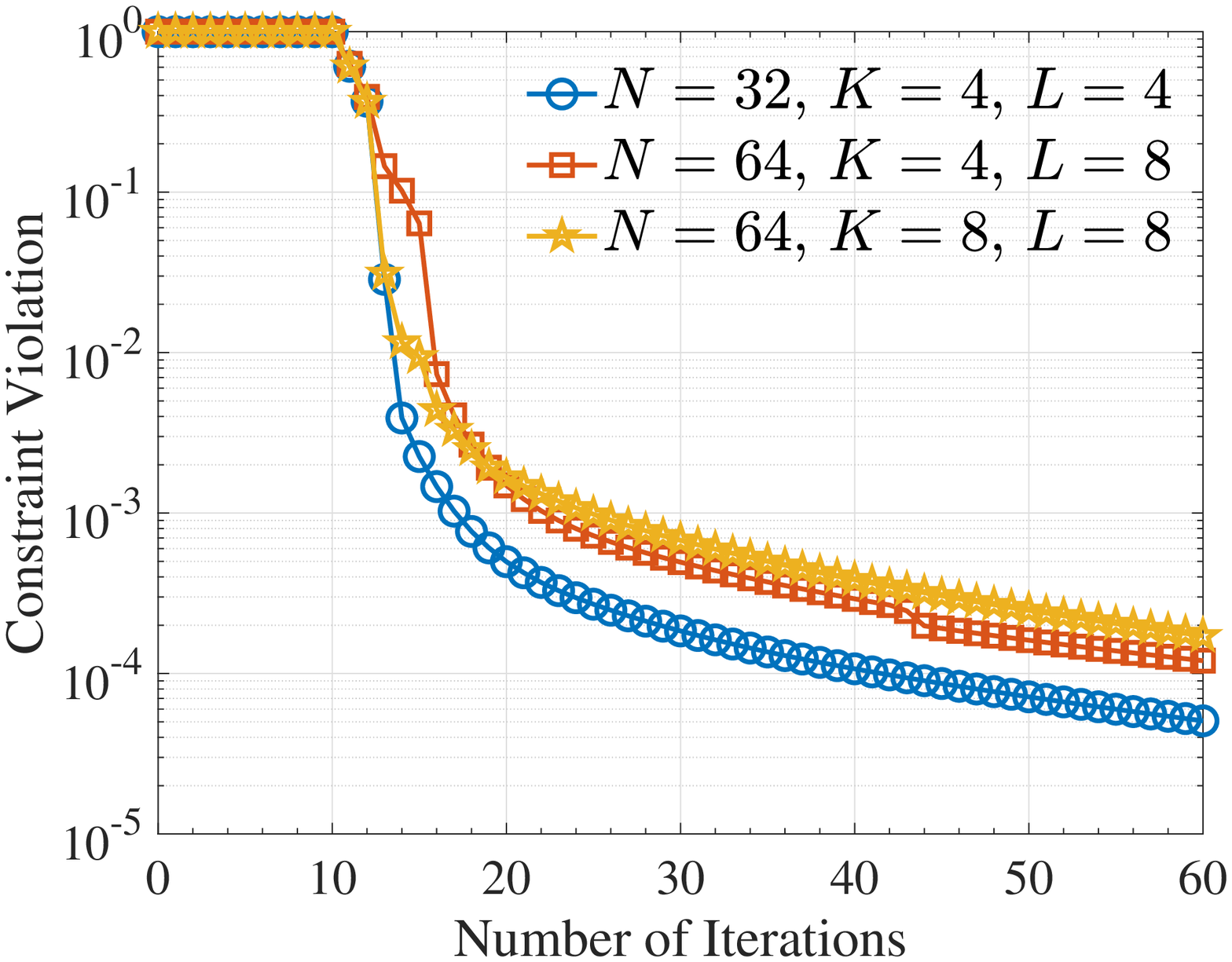}
	   \label{fig1b}	
    }
\caption{Average convergence performances ($p_{\max}=5$ dBm).}
    \label{figure1}
    \vspace{-20pt}
\end{figure}

\begin{figure*}[!t]
    \centering
    \subfigbottomskip=0pt
	\subfigcapskip=-5pt
\setlength{\abovecaptionskip}{0pt}
   \subfigure[Sum rate versus the transmit power.]
    {
        \includegraphics[height=0.23\textwidth]{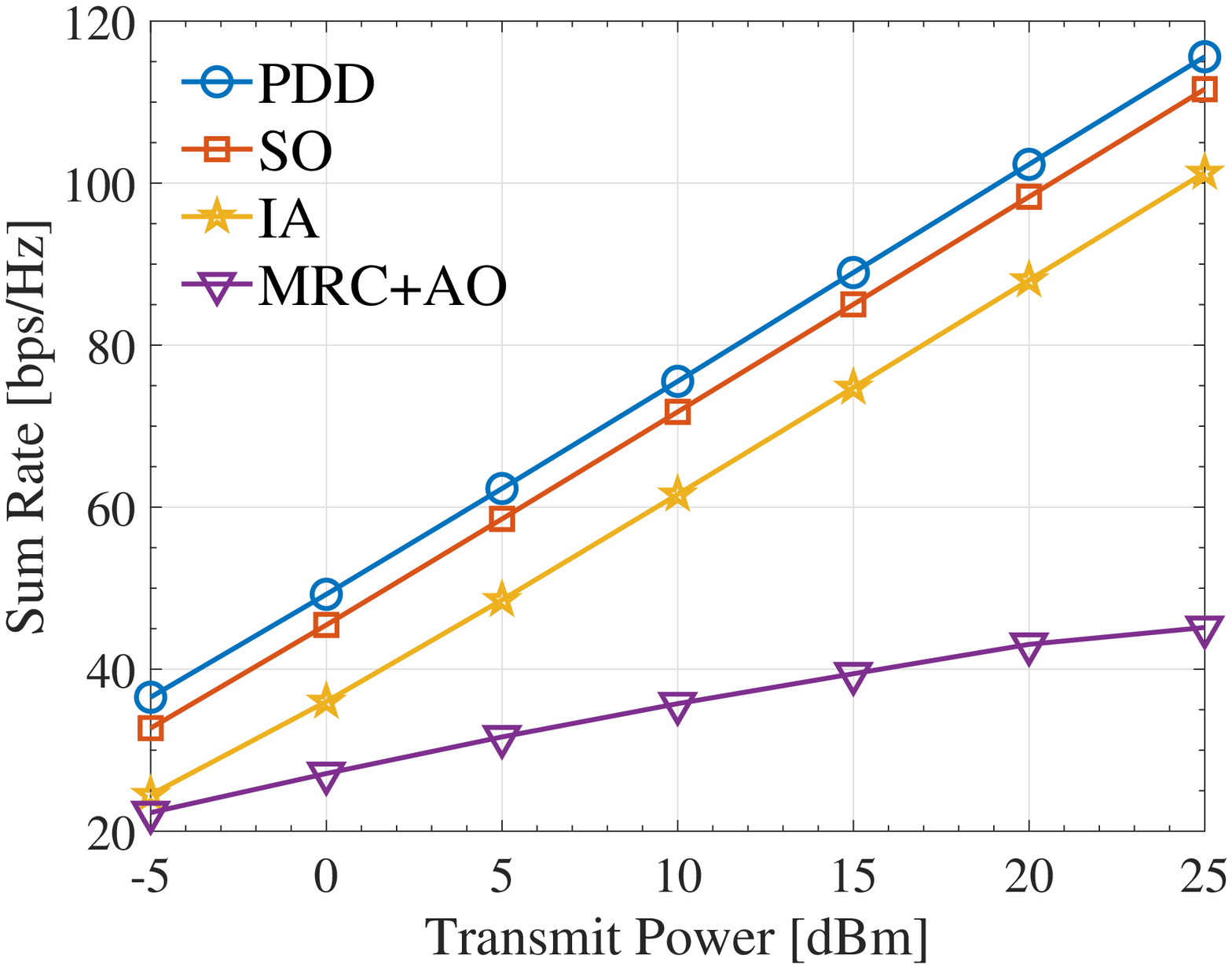}
	   \label{fig2a}	
    }
    \subfigure[Sum rate versus the UT antenna number.]
    {
        \includegraphics[height=0.23\textwidth]{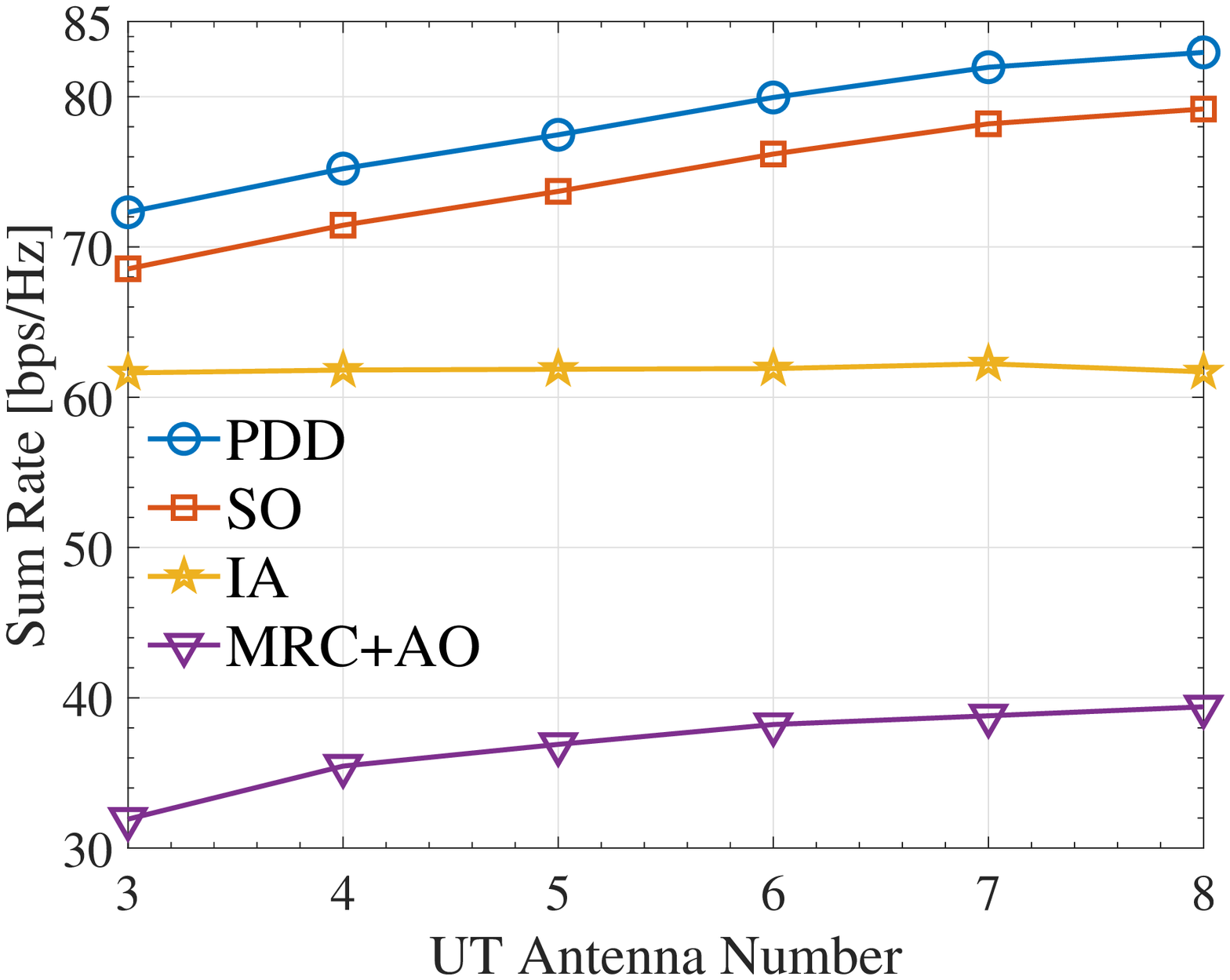}
	   \label{fig2b}	
    }
    \subfigure[Sum rate versus the RF chain number.]
    {
        \includegraphics[height=0.23\textwidth]{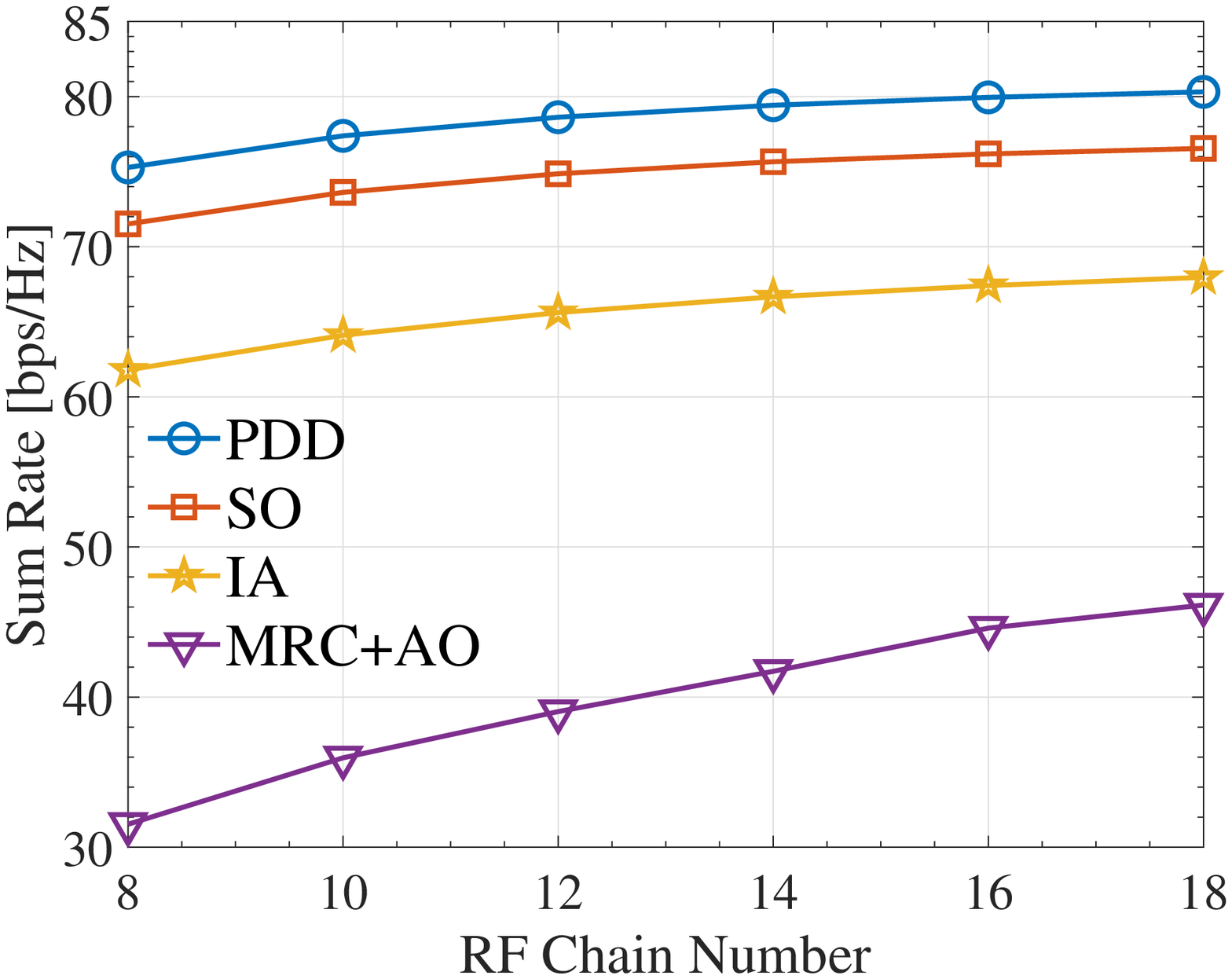}
	   \label{fig2c}	
    }
   \caption{Sum rate performances of the proposed algorithms. The simulation parameters are given as follows: (a) $L=8$ and $N_k=4$ ($\forall k$); (b) $L=8$ and $p_{\max}=10$ dBm; (c) $N_k=4$ ($\forall k$) and $p_{\max}=10$ dBm.}
    \label{figure2}
    \vspace{-20pt}
\end{figure*}

\section{Numerical Results}\vspace{-5pt}
Unless stated otherwise, the following simulation parameters are used: $N=64$, $N_k=N_{\rm{ut}}$, $p_{k,\max}=p_{\max}$, $N_{k}=4$, $M_k=4$, $\forall k$, $K=8$, and $\sigma^2=-100$ dBm. All the angles in the considered S-V model are randomly set within $[0, 2\pi)$. Furthermore, the UTs are assumed to be distributed uniformly over a hexagonal cell with radius of 100 m. The free-space path loss model is $-10\log_{10}\mu_k=92.5+20\log_{10}\left[f_0\left(\text{GHz}\right)\right]+20\log_{10}\left[d_k\left(\text{km}\right)\right]$, where $f_0=28$ GHz is the carrier frequency and $d_k$ is the distance between the BS and UT $k$. We comment that all the optimization variables are randomly initialized.

At first, we use {\figurename} {\ref{figure1}} to illustrate the average convergence performances of the proposed PDD-based method. Specifically, as can be seen from {\figurename} {\ref{fig1a}}, the achievable system sum rate converges rapidly in less than 15 outer iterations. Furthermore, as shown in {\figurename} {\ref{fig1b}}, the constraint violation $h$ reduces to a threshold $\mu=10^{-3}$ in less than 30 outer iterations, which means that the solution has essentially met the equality constraints for problem $\mathcal{P}_1$.

In {\figurename} {\ref{figure2}}, the sum rate performances achieved by the following schemes are presented for comparison:
\begin{enumerate}
  \item PDD: This is our proposed PDD-based method.
  \item SO: This is the sequential optimization-based method.
  \item IA: This is a baseline scheme where the phase shifts are designed randomly and the beam selection matrix is designed by the classical IA scheme \cite{Gao2016,Tataria2018,Guo2018,Feng2019,Liu2021}.
  \item MRC-AO: This is another baseline scheme proposed in \cite{Cheng2022_2}, where an MRC strategy is adopted for signal detection. The phase-only beamforming at the UTs and the beam selection at the BS are jointly designed by an altenating optimization (AO)-based scheme.
\end{enumerate}

Particularly, {\figurename} {\ref{fig2a}} plots the SR performances of the above four schemes versus the transmit power $p_{\max}$. It can be seen from this graph that the proposed PDD-based and SO-based methods can achieve higher sum rates than the baseline IA and MRC-AO schemes. It is worth mentioning that the SO-based method can yield a very close sum rate performance to that of the PDD-based method. This, together with the fact that the SO-based method involves lower computational complexity, suggests that the SO-based scheme is more preferred in practical systems. Moreover, the computational complexity of the baseline MRC-AO method is given by $\mathcal{O}\left(I_1\left(N_{\rm{U}}^{1.5}+I_2I_3N^{3.5}\right)\right)$ with $I_1$, $I_2$, and $I_3$ denoting numbers of iterations \cite{Cheng2022_2}. Clearly, the complexity of the MRC-AO method is approximately the same as that of the PDD-based method while is much higher than that of the SO-based method. Since the PDD-based and SO-based methods can yield higher sum rates, it can be concluded that our proposed method is superior to the state-of-the-art MRC-AO method. In {\figurename} {\ref{fig2b}}, we plot the sum rate as a function of the UT antenna number $N_{\rm{ut}}$. As can be observed, the sum rate achieved by the IA-based method enjoys few improvements when $N_{\rm{u}}$ increases. The reason lies in that the phase-only beamforming vectors therein are randomly set. In contrast to this scheme, the sum rate achieved by the other three schemes increase with $N_{\rm{ut}}$, which highlights the significance of phase shifts or phase-only beamforming design. Finally, we show the sum rate versus the number of RF chains in {\figurename} {\ref{fig2c}}. As expected, the proposed two schemes can outperform the baseline schemes in terms of the sum rate performance. Moreover, it can be seen from this figure that the sum rate can get improved by increasing the number of RF chains. Taken together, it can be found that the proposed joint design algorithms can effectively improve the system throughput.

\section{Conclusion}
The LAA-assisted MU-MIMO uplink mmWave communication systems have been investigated under the sum rate maximization criterion. An efficient joint phase-only beamforming and beam selection design algorithm has been formulated based on the PDD method. To reduce the computational complexity, a simplified sequential optimization-based method has also been presented. Numerical results have verified the superiority of the proposed methods over existing baselines.

\end{document}